\documentclass[12pt,preprint]{aastex}

\shorttitle{The Parker Instability}
\shortauthors{Kim et al.}
\slugcomment{draft of \today}

\begin{document}

\def\etal{{\it et al.~}}
\def\eg{{\it e.g.,~}}
\def\ie{{\it i.e.,~}}

\title{The Effect of the Random Magnetic Field Component\\
on the Parker Instability}

\author{Jongsoo Kim}

\affil{Korea Astronomy Observatory, 61-1, Hwaam-Dong, Yusong-Ku,
       Taejon 305-348, Korea \& NCSA, University of Illinois at
       Urbana-Champaign, 405 North Mathews Avenue, Urbana, IL 61801:
       jskim@ncsa.uiuc.edu}

\and

\author{Dongsu Ryu}

\affil{Department of Astronomy \& Space Science, Chungnam National
       University, Daejeon 705-764, Korea: ryu@canopus.chungnam.ac.kr}

\begin{abstract}

The Parker instability is considered to play important roles
in the evolution of the interstellar medium.
Most studies on the development of the instability so far have been
based on an initial equilibrium system with a uniform magnetic field.
However, the Galactic magnetic field possesses a random component
in addition to the mean uniform component, with comparable strength
of the two components. Parker and Jokipii have recently suggested
that the random component can suppress the growth of small wavelength
perturbations. Here, we extend their analysis by including gas
pressure which was ignored in their work, and study the stabilizing
effect of the random component in the interstellar gas with finite
pressure. Following Parker and Jokipii, the magnetic field is modeled
as a mean azimuthal component, $B(z)$, plus a random radial
component, $\epsilon(z) B(z)$, where $\epsilon(z)$ is a random
function of height from the equatorial plane.
We show that for the observationally suggested
values of $\langle\epsilon^2\rangle^{1/2}$, the tension due to the
random component becomes important, so that the growth of the
instability is either significantly reduced or completely suppressed.
When the instability still works, the radial wavenumber of the most
unstable mode is found to be zero. That is, the instability is reduced
to be effectively two-dimensional. We discuss briefly the implications
of our finding.

\end{abstract}

\keywords{instabilities --- ISM: clouds --- ISM: magnetic fields ---
magnetohydrodynamics: MHD}

\section{Introduction}

The magnetostatic equilibrium of the system of the interstellar gas
and magnetic field under the vertical gravitational field of the
Galaxy has been shown to be unstable \citep{par66,par67}.
The physical mechanism for the instability relies on the fact that
a light fluid (represented by the magnetic field) supports a heavy
fluid (represented by the gas) and the configuration tends to overturn.
It has similarities to the Rayleigh-Taylor instability, when a true
light fluid supports a heavy fluid. In the Rayleigh-Taylor instability,
the fastest growing mode has an infinite perturbation wavenumber.
However, taking into account a ``uni-directional'' magnetic
field along the azimuthal direction in the Galactic disk,
\citet{par66} showed that the magnetic tension
stabilizes large wavenumber perturbations and results in a preferred,
finite wavenumber. But when the
perturbations along the radial direction are allowed, those
with an infinite radial wavenumber prevail \citep{par67}.
As a result, the structures formed by the Parker instability
are expected to be elongated, and \citet{kim98} confirmed it
through three-dimensional simulations for the nonlinear evolution of
the Parker instability.

The Parker instability in the interstellar medium (ISM) has been
thought to be a viable mechanism in forming giant molecular clouds
(GMCs) in the Galaxy \citep[see, \eg][]{app74,mou74,bli80}.
However, the work of \citet{kim98} raised negative points on that.
In addition to the fact that sheet-like structures with the smallest
scale in the radial direction are formed, they found that the
enhancement factor of column density is at most $\sim2$. The
second density issue is eased by noting that the interstellar gas
can be further susceptible to the thermal instability, as pointed
in \citet{par00}, followed by the gravitational instability.
However, the first structural issue, which is the direct result of
the infinitesimal radial wavenumber, does not easily go away.

Several ideas on effects that could suppress the maximally unstable
nature of the mode with an infinite wavenumber have been suggested.
One of them is to invoke a ``stochastic magnetic field'' \citep{par00},
which represents the random component of the Galactic magnetic field.
Using a field composed of the usual mean component and a transverse
component whose strength is weak and random, they showed
that in ``cold plasma'' (without gas pressure) the weak, random
component exerts a significant stabilizing effect on the perturbations
with small transverse wavelengths. The physical mechanism is the
following. Although weak, the tension of the transverse component
that is incurred by the vertical gas motions
is strong enough to reinstate the gas. They suggested the possibility
of preferred modes with finite transverse (radial) wavenumbers.
Such modes would result in broadened structures, which would resemble
more the morphology of the GMCs.
Their stochastic field model is promising in the sense that i) it is
consistent with the turbulent picture of the ISM
\citep[see, \eg][]{ms96} and ii) it is supported by the observations
of magnetic field in our Galaxy and spiral galaxies
\citep[see, \eg][]{bbmss96,zh97}.
However, their cold plasma approximation needs to be improved.

The purpose of this paper is to analyze fully the effects of the
random component of the magnetic field on the Parker
instability in a medium with finite gas pressure. We find rather
surprising results that the random component either reduces
the growth of the instability significantly or suppresses it completely.
And the most unstable mode has a vanishing radial wavenumber.
The plan of the paper is as follows.  Linear stability
analysis is carried out by analyzing the dispersion relation
in \S 2. Summary and discussion follow in \S 3.

\section{Linear Stability Analysis}

We consider the stability of an equilibrium system where gas is supported
by its own and magnetic pressures against a ``uniform'' gravity, $g$,
in the negative
$z$ (vertical) direction. With realistic gravities different growth rates
and wavelengths of unstable modes would result \citep[see, \eg][]{khr97},
but they make the analysis much more involved.
In addition, we expect the qualitative features of the stability
wouldn't be affected by details of gravity. For the magnetic field
configuration, the stochastic model suggested by \citet{par00} is
adopted. It is composed of a mean component, $B(z)$, in the $y$
(azimuthal) direction, and a random component, $\epsilon(z)B(z)$, in the $x$
(radial) direction. $\epsilon(z)$ is a random function of $z$ with zero mean.
One assumption made on $\epsilon(z)$ is that the correlation length
is small compared to the vertical scale height of the system.
So in the equations below, the local
average is taken by integrating over $z$ for a vertical scale
greater than the correlation length of $\epsilon(z)$ but smaller than
the scale height. Then, the dispersion $\langle\epsilon^2\rangle$ is taken
as a constant, which becomes a free parameter of the analysis.
With finite gas pressure, $p$, the magnetohydrostatic equilibrium is
governed by
\begin{equation}
\frac{d}{dz}\left[ p + (1+\langle\epsilon^2\rangle)\frac{B^2}{8\pi} \right]
= - \rho g,
\end{equation}
where $\rho$ is gas density. Two further assumptions are made, which are
usual in the analysis of the Parker instability: i) an isothermal
equation of state, $p = a_s^2 \rho$, where $a_s$ is the isothermal speed,
and ii) a constant ratio  of magnetic to gas pressures,
$\alpha=(1+\langle\epsilon^2\rangle)B^2/(8\pi p)$.
Then exponential distributions of density, gas pressure, and magnetic
pressure are obtained
\begin{equation}
\frac{\rho(z)}{\rho(0)} =
\frac{p(z)}{p(0)} =
\frac{B^2(z)}{B^2(0)} =
\exp \left( - \frac{|z|}{H} \right)_,
\end{equation}
where the e-folding scale height, $H$, is given by $(1+\alpha)a_s^2/g$.

The above equilibrium state is disturbed with an infinitesimal perturbation.
The perturbed system is assumed to be isothermal too. Since linearized
perturbation equations for the case without gas pressure were
already derived \citep{par00}, the detailed derivation is not
repeated here. Instead, a reduced form in terms of velocity
perturbations, $(v_x,v_y,v_z)$, is written down as follows:
\begin{eqnarray} \label{velx}
\frac{\partial^2 v_x}{\partial t^2} &=&
a_s^2 \frac{\partial}{\partial x}
      \left( \frac{\partial v_x}{\partial x}
            +\frac{\partial v_y}{\partial y}
            +\frac{\partial v_z}{\partial z}
            -\frac{v_z}{H} \right) \nonumber \\
&+& v_A^2 \left[ \frac{\partial^2 v_x}{\partial y^2}
            +\frac{\partial^2 v_x}{\partial x^2}
            +\frac{\partial}{\partial x}
                  \left( \frac{\partial v_z}{\partial z}
                        -\frac{v_z}{2H} \right)
            -\frac{\langle\epsilon^2\rangle}{2H}\frac{\partial v_z}{\partial x}
      \right]_,
\end{eqnarray}
\begin{eqnarray} \label{vely}
\frac{\partial^2 v_y}{\partial t^2} &=&
a_s^2 \frac{\partial}{\partial y}
      \left( \frac{\partial v_x}{\partial x}
            +\frac{\partial v_y}{\partial y}
            +\frac{\partial v_z}{\partial z}
            -\frac{v_z}{H} \right) \nonumber \\
&+& v_A^2 \left\{-\frac{1}{2H}\frac{\partial v_z}{\partial y}
       +\langle\epsilon^2\rangle
           \left[ \frac{\partial^2 v_y}{\partial x^2}
                 +\frac{\partial^2 v_y}{\partial y^2}
                 +\frac{\partial}{\partial y}
                      \left( \frac{\partial v_z}{\partial z}
                            -\frac{v_z}{2H}
                      \right)
           \right]
      \right\}_,
\end{eqnarray}
\begin{eqnarray} \label{velz}
\frac{\partial^2 v_z}{\partial t^2} &=&
a_s^2 \frac{\partial}{\partial z}
      \left( \frac{\partial v_x}{\partial x}
            +\frac{\partial v_y}{\partial y}
            +\frac{\partial v_z}{\partial z}
            -\frac{v_z}{H} \right) \nonumber \\
&+& v_A^2 \left\{ \frac{\partial^2 v_z}{\partial y^2}
                 +\left( \frac{\partial}{\partial z}
                        -\frac{1}{H} \right)
                  \left[ \left( \frac{\partial}{\partial z}
                               -\frac{1}{2H} \right) v_z
                        +\frac{\partial v_x}{\partial x}
                  \right] \right. \nonumber \\
                 &&\left. + \frac{1}{2H}
                       \left( \frac{\partial v_x}{\partial x}
                      +\frac{\partial v_y}{\partial y}
                      +\frac{\partial v_z}{\partial z}
                      -\frac{v_z}{H} \right) \right. \nonumber \\
                 &&\left. + \langle\epsilon^2\rangle
                   \left[\left( \frac{\partial}{\partial z}
                               -\frac{1}{H} \right)
                         \left( \frac{\partial}{\partial z}
                               -\frac{1}{2H} \right) v_z
                        +\left( \frac{\partial}{\partial z}
                               -\frac{1}{H} \right)
                         \frac{\partial v_y}{\partial y}
                        +\frac{\partial^2 v_z}{\partial x^2}
                   \right] \right\}_.
\end{eqnarray}
Here, $v_A$ is the Alfv\'en speed, $B/\sqrt{4\pi\rho}$,
which is constant over $z$. Note that the linearized perturbation
equations for the cold plasma (Eqs. [12] - [14] in \citet{par00}) are
recovered from the above equations by i) dropping out the terms with $a_s$
and ii) noting that the scale height of magnetic field ($\Lambda$ of their
notation) is twice larger than that of gas ($H$ of our notation).

The normal mode solution takes the following form
\begin{equation} \label{sol}
(v_x,v_y,v_z) = (D_x,D_y,D_z)
              \exp \left(
                   \frac{t}{\tau} + ik_x x + ik_y y + ik_z z + \frac{z}{2H}
                   \right)_,
\end{equation}
where $D_x$, $D_y$, and $D_z$ are constants. Taking $H$ and $H/a_s$ as
the normalization units of length and time, respectively, the dimensionless
growth rate, $\Omega=H/(a_s\tau)$, and the dimensionless wavenumber,
$(q_x,q_y,q_z) = H (k_x,k_y,k_z)$, are defined.  Substituting
Eq. (\ref{sol}) into Eqs. (\ref{velx})-(\ref{velz}) and imposing the
condition of a non-trivial solution, we get the dispersion relation
\begin{equation} \label{disprel}
\Omega^6 + C_4 \Omega^4 + C_2 \Omega^2 + C_0 = 0,
\end{equation}
where the coefficients $C_4$, $C_2$ and $C_0$ are given by
\begin{equation}
C_4 = 2\alpha q_y^2 + (2\alpha+1)(q_x^2+q_y^2+q_z^2+1/4)
    + 2\alpha\langle\epsilon^2\rangle(2q_x^2+q_y^2+q_z^2+iq_z/2),
\end{equation}
\begin{eqnarray}
C_2 &=& \alpha(\alpha+1) \left[ q_x^2 + 4q_y^2(q_x^2+q_y^2+q_z^2)
                         \right] \nonumber \\
    &+& \alpha\langle\epsilon^2\rangle\left\{\alpha q_x^2 + (2\alpha+1)q_y^2
        +4q_x^2\left[(2\alpha+1)(q_x^2+q_z^2)+(4\alpha+1)q_y^2 \right]
        + 4\alpha q_y^2 \left[ 2(q_y^2+q_z^2) + iq_z/2 \right]
        \right\} \nonumber \\
    &+& 4\alpha^2\langle\epsilon^2\rangle^2 q_x^2 (q_x^2+q_y^2+q_z^2+iq_z /2),
\end{eqnarray}
\begin{eqnarray}
C_0 &=& 2\alpha^2 q_y^2 \left[ 2q_y^2(q_x^2+q_y^2+q_z^2+1/4)
                              -(\alpha+1)(q_x^2+q_y^2) \right] \nonumber \\
    &+& 2\alpha^2\langle\epsilon^2\rangle
        \left\{
               \left[ (\alpha+1)q_x^4+3(\alpha+1)q_x^2q_y^2 +(2\alpha+1)q_y^4
               \right]
               +4q_y^2 (q_x^2+q_y^2+q_z^2) \left[ q_x^2+\alpha(q_x^2+q_y^2)
                                           \right]
        \right\} \nonumber \\
    &-& \alpha^2\langle\epsilon^2\rangle^2 q_x^2
        \left\{ q_x^2+2q_y^2 - 4(q_x^2+q_y^2+q_z^2)
                               \left[q_x^2+2\alpha(q_x^2+q_y^2)
                               \right]
        \right\}.
\end{eqnarray}
Eq. (\ref{disprel}) is a cubic equation of $\Omega^2$ with complex
coefficients. For the case with vanishing vertical wavenumber
($q_z=0$), all the $C$ coefficients become real, and the dispersion
relation can be easily solved. For small vertical wavenumbers,
the imaginary terms, $i\langle\epsilon^2\rangle q_z$ and
$i\langle\epsilon^2\rangle^2 q_z$
in $C_4$ and $C_2$, can be still ignored. This trick doesn't affect
the marginal condition of the stability ($\Omega=0$),
since $C_0$ doesn't contain any imaginary term.
Here, we remind readers of the definition of $\alpha$.  It is reserved in
this paper for the ratio of magnetic to gas pressures, whereas it was used
for the dispersion of $\epsilon$ in \citet{par00}.  For the dispersion
$\langle\epsilon^2\rangle$ is used in this paper.

Two limiting cases can be considered, which enable us to check the validity
of the above relation. The formula with $\langle\epsilon^2\rangle=0$
reduces to the
dispersion relation for the original Parker instability
\citep[see, \eg][]{par67,Shu74}. The $C$'s without the terms containing
$\langle\epsilon^2\rangle$ match exactly with the coefficients of Eq.~(53) in
\citep{Shu74}, after imposing the isothermal condition $\gamma=1$.
The other limiting formula is for the cold plasma with $p=0$ \citep{par00}.
As shown above, our linearized perturbation equations recover those
for the cold plasma.

The full stability property can be analyzed by solving the above dispersion
relation numerically. Fig. 1 shows the stability diagram for $\alpha=1$.
Equi-$\Omega^2$ contours with positive values corresponding to unstable modes
are plotted on the $(q_x^2,~q_y^2)$ plane for a few different values
of $\langle\epsilon^2\rangle^{1/2}$. $q_z=0$ has been set.
Finite $q_z$'s reduce the growth rate \citep[see, \eg][]{par66}.
Three interesting points can be made: with increasing
strength of the random component, i) the domain of the instability in
the $(q_x^2,~q_y^2)$ plane shrinks, ii) the maximum growth rate decreases,
and iii) the $q_{x,{\rm max}}$, which gives the maximum growth rate, decreases
and reduces to zero eventually. Note that without
the random component, $\langle\epsilon^2\rangle^{1/2}=0$,
the most unstable mode has
the growth rate $\Omega^2=0.172$ and the radial wavenumber
$q_{x,{\rm max}} \rightarrow \infty$.

The above points can be seen more clearly in Fig. 2, which shows
the growth rate and two horizontal wavenumbers of the most unstable
modes as a function of $\langle\epsilon^2\rangle^{1/2}$ for three
different values of $\alpha$. Again, $q_z=0$ has been set.
Note that the scale height $H$ changes with $\alpha$.
Hence, both the growth rate and wavenumber in real units
scale as $1/(1+\alpha)$. Even after this factor is taken into account,
the maximum growth rate increases with $\alpha$,
due to enhanced magnetic buoyancy. Two additional points can be made:
i) the critical value $\langle\epsilon^2\rangle_c^{1/2}$,
above which the Parker instability disappears completely,
is independent of $\alpha$, and it is computed as $1/\sqrt{2}=0.707$
from the dispersion relation, and ii) the value of
$\langle\epsilon^2\rangle^{1/2}$, above
which the radial wavenumber of the most unstable mode vanishes, decreases
with increasing $\alpha$. These are the consequences of different roles
of uniform and random magnetic fields.

On the issue of the formation of GMCs, the most interesting
range of the values would be $\langle\epsilon^2\rangle^{1/2} \sim 0.1$,
which would result in $q_x \sim q_y$ for the most unstable modes
(although it would depend on $\alpha$, see Fig. 2).
Then, the structures formed
as the result of the instability would be round, so mimicking GMCs.
However, observations suggest a larger random component. It is
generally quoted that in the Galactic plane
$0.5\la\langle\epsilon^2\rangle^{1/2}\la1$,
with the strength of the total magnetic field $B\sim 3-4 \mu{\rm G}$
\citep[see, \eg][]{bbmss96,zh97}. But others such as \citet{ms96} suggest
somewhat smaller values such as
$\langle\epsilon^2\rangle^{1/2} \sim 1/4 - 1/3$.
If $0.5 \la \langle\epsilon^2\rangle^{1/2} \la 1$, the instability disappears
completely or the growth rate reduces significantly by more than 80\%.
If $\langle\epsilon^2\rangle^{1/2} \sim 1/4 - 1/3$, the growth rate reduces by
$30 - 60$\%. But in any case the radial wavenumber of the most unstable mode
shrinks to zero $(q_x =0)$. {\it That is, the instability basically
becomes two-dimensional in the plane defined by the azimuthal and
vertical directions.} This result is opposite to that of
$\langle\epsilon^2\rangle^{1/2}=0$, where the dominant mode of the instability
has vanishing radial wavelength $(q_x \rightarrow \infty)$.

\section{Discussion and Summary}

The Parker instability is induced by the magnetic buoyancy of uniform
component of magnetic field, while gas pressure and random component
exert stabilizing effects. The role of gas pressure is mainly
exercising pressure force along the uniform component, and set finite
wavenumbers for the instability along the uniform component
direction. On the other hand,
random component threads rising and sinking slices across
the uniform field. Its tension becomes stronger at larger wavenumbers.
So the role of the random field is to suppress the growth of
perturbations with large wavenumbers perpendicular to the uniform field.

Through the linear stability analysis which includes both
gas pressure and random magnetic field,
we have found that with the observationally favored values for
the strength of the random component,
$0.5 \la \langle\epsilon^2\rangle^{1/2} \la 1$,
the tension of the random component that is incurred
by the vertical gas motions is strong enough that the growth of the
instability is either significantly reduced or completely suppressed.
For smaller values, $\langle\epsilon^2\rangle^{1/2} \sim 1/4 - 1/3$, which are
suggested by others, the Parker instability is still operating
but with reduced growth rate and vanishing radial wavenumber.
With $\langle\epsilon^2\rangle^{1/2} \sim 1/4 - 1/3$,
by taking $H=160$ pc and $a_s=6.4$ km/s
\citep{fl73}, the growth time scale and the azimuthal wavelength of the
most unstable mode are $70-95$ Myrs and $\sim 2.2$ kpc, respectively.
They are too large for the Parker instability to be a plausible mechanism
for the formation of GMCs. But it is known that realistic gravity
would reduce both \citep[see, \eg][]{khr97}. The more serious obstacle
in the context of the GMC formation is the fact that the radial
wavelength of the most unstable mode is infinity. This indicates the
structures formed would be elongated, in this case, along the radial
direction. But it is not clear whether such elongated structures would
persist in the stage of the nonlinear development of the instability.
That should be tested by numerical simulations.

\acknowledgments

The work was supported in part by KRF through grant KRF-2000-015-DS0046.
We thank Dr. R. Jokipii for discussions and Dr. T. W. Jones for comments
on the manuscript.

\begin{figure}
\epsscale{0.3562}\plotone{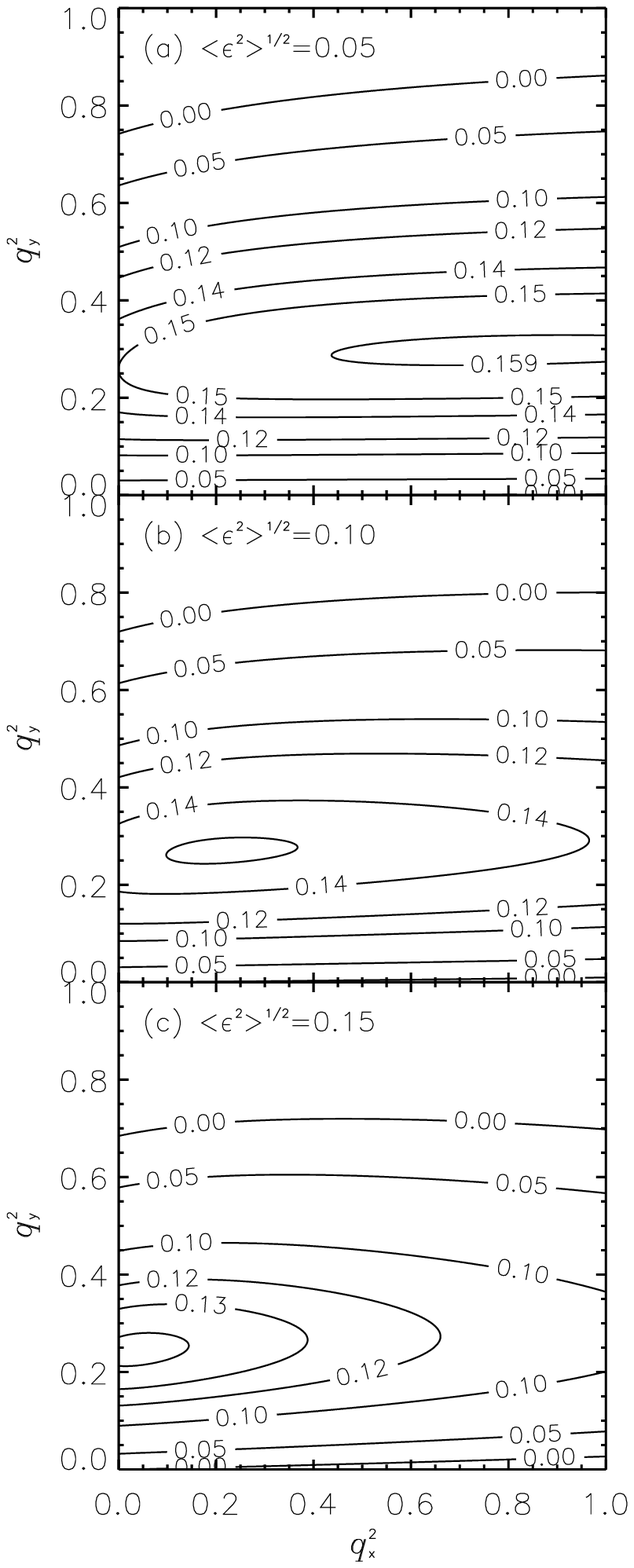}\epsscale{1.}
\figcaption{Equi-$\Omega^2$ contours on the $(q_x^2,q_y^2)$ plane
for (a) $\langle\epsilon^2\rangle^{1/2}$ = 0.05, 
(b) $\langle\epsilon^2\rangle^{1/2}$ =0.10, 
(c) $\langle\epsilon^2\rangle^{1/2}$ =0.15.
In all plots, $\alpha=1$ and $q_z=0$ are used.\label{fig1}}
\end{figure}

\begin{figure}
\plotone{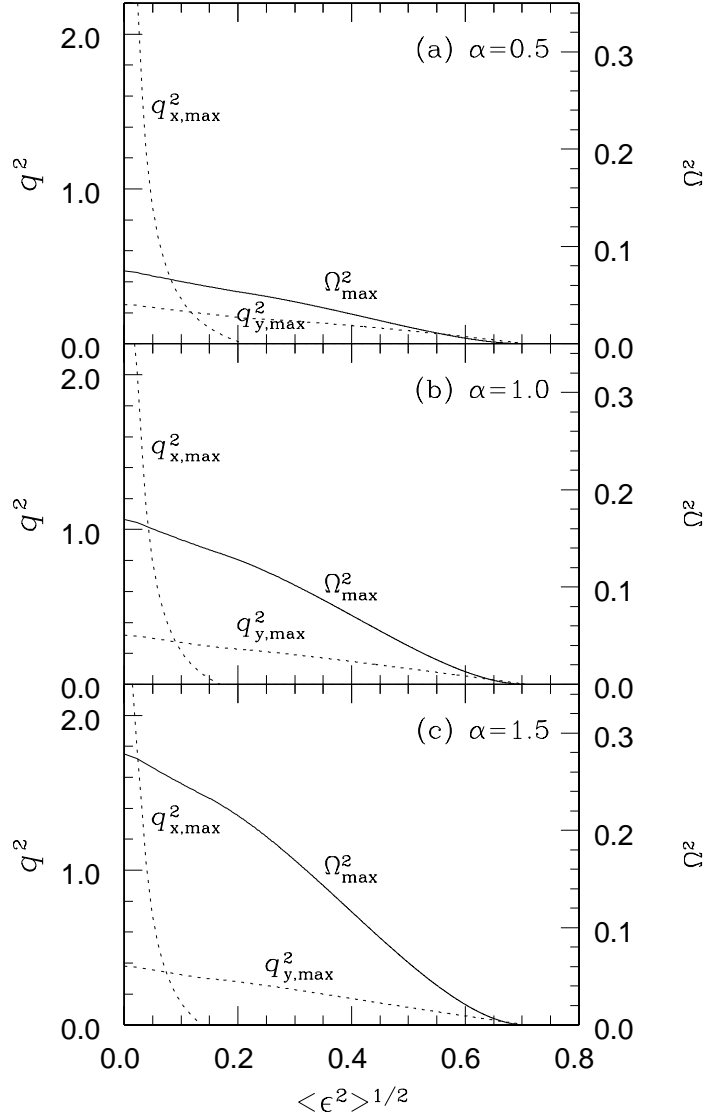}
\figcaption{Maximum growth rate, $\Omega^2_{\rm max}$, and its two 
horizontal wavenumbers, $q^2_{x,\rm max}$ and $q^2_{y,\rm max}$, 
as a function of $\langle\epsilon^2\rangle^{1/2}$ for (a) $\alpha$=0.5, 
(b) $\alpha=1.0$ and (c) $\alpha=1.5$. $q_z=0$ for all the cases.
Note that the growth rate and wavenumber in real units
scale as $1/(1+\alpha)$.\label{fig2}}
\end{figure}


\begin{thebibliography}{}

\bibitem[Appenzeller(1974)]{app74}
Appenzeller, I. 1974, \aap, 36, 99
\bibitem[Beck \etal(1996)]{bbmss96}
Beck, R., Brandenburg, A., Moss, D., Shukurov, A., \& Sokoloff, D.
1996, \araa, 34, 155
\bibitem[Blitz \& Shu(1980)]{bli80}
Blitz, L., \& Shu, F. H. 1980, \apj, 238, 148
\bibitem[Falgarone \& Lequeux(1973)]{fl73}
Falgarone, E., \& Lequeux, J. 1973, \aap, 25, 253.
\bibitem[Kim \etal(1997)]{khr97}
Kim, J., Hong, S. S., \& Ryu, D.,1997, \apj, 485, 228
\bibitem[Kim \etal(1998)]{kim98}
Kim, J., Hong, S. S., Ryu, D., \& Jones, T. W. 1998, \apj, 506, L139
\bibitem[Minter \& Spangler(1996)]{ms96}
Minter, A. H., \& Spangler, S. R., 1996, \apj, 194, 214
\bibitem[Mouschovias \etal(1974)]{mou74}
Mouschovias, T. Ch., Shu, F. H., \& Woodward, P. R. 1974, \aap, 33, 73
\bibitem[Parker(1966)]{par66}
Parker, E. N. 1966, \apj, 145, 811
\bibitem[Parker(1967)]{par67}
Parker, E. N. 1967, \apj, 149, 535
\bibitem[Parker \& Jopikii(2000)]{par00}
Parker, E. N., \& Jokipii, J. R. 2000, \apj, 536, 331
\bibitem[Shu(1974)]{Shu74}
Shu, F. H. 1974, \aap, 33, 55
\bibitem[Zweibel \& Heiles(1997)]{zh97}
Zweibel, E. G. \& Heiles, C., 1997, Nature, 385, 131

\end{thebibliography}
\end{document}